\documentclass[reprint,superscriptaddress,amsmath,amssymb]{revtex4-1}

\usepackage{graphicx,bm,hyperref,natbib,color}

\newcommand{\figref}[1]{Fig.\ \ref{#1}}
\newcommand{\textdensity}[2]{#1$\times10^{#2}\text{ cm}^{-2}$}
\newcommand{\textmobility}[1]{#1 cm$^2$/V$\cdot$s}
\newcommand{\fullTaWSi}{Ta$_{40}$W$_{40}$Si$_{20}$}
\newcommand{\fullALD}{Al$_2$O$_3$}
\newcommand{\textmicron}{$\mu$m}

\begin{document}
\title{Transport Measurements of Surface Electrons in 200 nm Deep Helium-Filled Microchannels Above Amorphous Metallic
Electrodes}
\author{A. T. Asfaw}
 \email{asfaw@princeton.edu}
\affiliation{
    Department of Electrical Engineering, Princeton University, Princeton, New Jersey 08544, USA
}
\author{E. I. Kleinbaum}
\affiliation{
    Department of Electrical Engineering, Princeton University, Princeton, New Jersey 08544, USA
}
\author{M. D. Henry}
\affiliation{
    Sandia National Laboratories, Albuquerque, New Mexico, USA
}
\author{E. A. Shaner}
\affiliation{
    Sandia National Laboratories, Albuquerque, New Mexico, USA
}
\author{S. A. Lyon}
\affiliation{
    Department of Electrical Engineering, Princeton University, Princeton, New Jersey 08544, USA
}
\date{\today}
\begin{abstract} 
    
    We report transport measurements of electrons on helium in a microchannel device where the channels are 200 nm
    deep and $3~\mu$m wide. The channels are fabricated above amorphous metallic \fullTaWSi, which has surface roughness
    below 1 nm and minimal variations in work function across the surface due to the absence of polycrystalline grains. We
    are able to set the electron density in the channels using a ground plane. We estimate a mobility
    of \textmobility{300} and electron densities as high as \textdensity{2.56}{9}. We demonstrate control of the transport
    using a barrier which enables pinchoff at a central microchannel connecting two reservoirs. The conductance
    through the central microchannel is measured to be 10 nS for an electron density of \textdensity{1.58}{9}. Our work
    extends transport measurements of surface electrons to thin helium films in microchannel devices above metallic substrates.

\end{abstract} 
\maketitle

For nearly five decades, the system of electrons floating on the surface of bulk helium has provided a physical platform
for studying a wide variety of phenomena\cite{andrei_two-dimensional_1997,monarkha_two-dimensional_2004}. The phase
diagram of this system can be further enriched at high electron densities where polaronic states, degenerate electron
fluids and superconductivity are predicted to exist \cite{peeters_electrons_1983,jiang_phase_1988}. This region of the
phase diagram is difficult to achieve due to two fundamental barriers. First, bulk helium exhibits a hydrodynamic
instability caused by the pressure of the electrons, which limits the maximum achievable electron density above the
helium\cite{etz_stability_1984,hu_stability_1990}. Second, because of the strong electron-electron interaction at high
densities, the Wigner crystal dominates the phase diagram for temperatures below 1 K\cite{grimes_crystallization_1980}.

These issues can be mitigated by using thin helium films on a dielectric substrate
\cite{etz_stability_1984,mistura_microwave_1997,jiang_phase_1988}.  The van der Waals interaction between the helium and
the substrate stiffens the film, pushing the hydrodynamic instability to higher electron densities. Further, the
presence of the dielectric introduces an image charge which screens the Coulomb interaction and reduces the area of the
Wigner crystal in the phase diagram.  With the use of substrates that have large dielectric constants, the shielding can
be enhanced. A metallic substrate would provide the highest density electron fluid.

A large body of experimental work has pursued these ends. While there is a significant body of work which has examined
transport of electrons above insulating substrates, there have been fewer studies of electron transport on metallic
substrates\cite{etz_stability_1984,angrik_electrons_2004}. The disparity between these two seemingly similar systems
arises from the presence of disorder in metallic substrates, which can easily suppress electron transport above the
metal and prevent basic transport measurements\cite{klier_electron_2008}. 

In this work, we fabricate and measure transport of electrons on a thin film of helium above an amorphous metallic substrate
in a microchannel device. In contrast with typical polycrystalline metals, amorphous metals are defined by the lack of
translational order on atomic length scales \cite{mcglonethesis}. As a consequence, the surfaces of thin-film amorphous metals are
exceptionally smooth and exhibit homogeneous work function over large areas, making them ideally suited for fabrication
of electrodes for electrons on helium. 

\begin{figure}
    \includegraphics[scale=0.30]{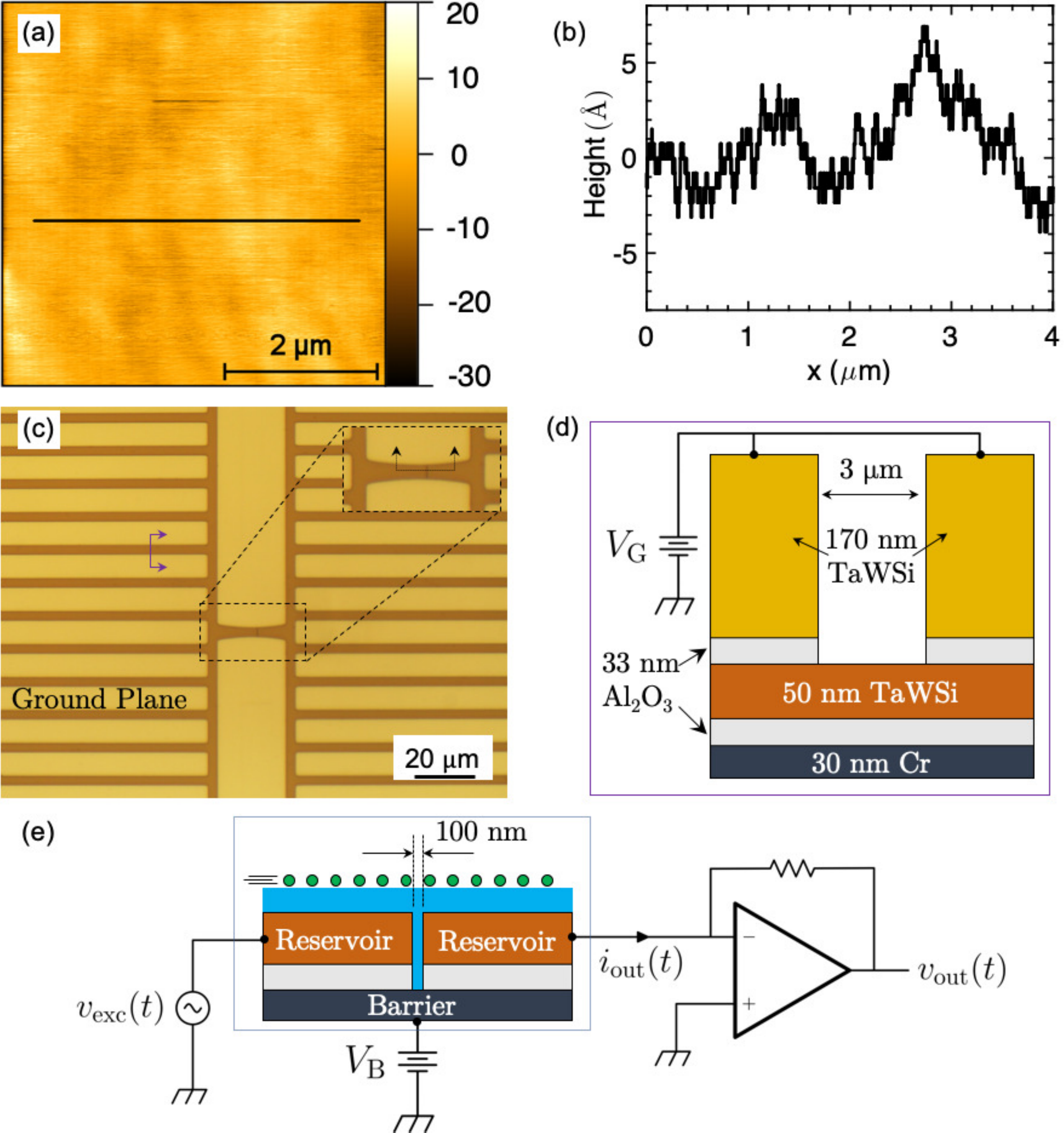}\caption{\label{fig1}(a) Topography of a 200 nm thick layer of
    RF-sputtered amorphous metallic \fullTaWSi\ measured using atomic force microscopy in a 5 \textmicron\ $\times$ 5
    \textmicron\ window. The colorbar is shown in units of Angstroms. The height profile along the indicated line in the
    middle of the scan is shown in (b). (c) Optical micrograph of the electrons on helium device. The channels are 200
    nm deep and 3 \textmicron\ wide. They are patterned on top of the left and right reservoir electrodes which are
    separated by a 100 nm gap as shown in the inset. (d) Cross-section of the device between the channels above the
    reservoir electrodes. (e) Cross-section of the device along the central microchannel. Figures (d) and (e) also show the
    electrical connections used throughout the measurements as described in the main text.}
\end{figure}

The amorphous metal used in this work is RF-sputtered \fullTaWSi\ (hereafter
TaWSi)\cite{mcglonethesis,mcglone_ta-based_2015,mcglone_tawsi_2017}. \figref{fig1}a shows the topography of a 200 nm thick layer of
TaWSi measured using atomic force microscopy in a 5 \textmicron\ $\times$ 5 \textmicron\ window, with a slice along the
indicated line in the middle of the scan shown in \figref{fig1}b. The RMS roughness is measured to be less than 5 $\AA$,
confirming the smoothness of the surface of the amorphous metal. In addition to these features, we have found that TaWSi is a
superconductor with a critical temperature near $2$ K. While this property of the metal has no noticeable effect on our measurements,
it provides a method of integrating superconducting electronics\cite{yang_coupling_2016}.

An optical micrograph of the device is shown in \figref{fig1}c. Electrons are confined to two sets of 3 \textmicron\
microchannels connected by a single central microchannel, similar to the device used by \citet{rees_transport_2012}. The
microchannels are filled by the capillary action of superfluid helium, which determines the depth of helium in the
channels.  The channels are defined in a 170 nm layer of TaWSi, referred to as the ground plane. Below the ground
plane is a 33 nm layer of insulating \fullALD\ yielding 200 nm deep channels. Additional electrodes are defined in
metallic layers beneath these channels. A 30 nm thick layer of chromium, referred to as the barrier, underlies the
entire device.  Above the barrier, the left and right reservoir electrodes are patterned in an additional 50 nm
thick layer of TaWSi. \figref{fig1}d and e show cross-sections of the device along the microchannels above the
reservoir electrodes and the central microchannel, respectively. The two reservoirs are separated by a 100 nm gap defined using
electron-beam lithography.

The device is placed in a leak-tight copper cell $0.5$ mm above the bulk helium level and cooled to a temperature of
1.7 K. Electrons are emitted onto the sample from a filament, with the ground plane voltage, $V_\text{G}$, set to -0.2 V and with all other
electrodes at 0 V. Subsequently, an excitation voltage, $v_{\text{exc}}(t) = 10\text{ mV}_{\text{RMS}}$ is applied to the left reservoir electrode at a
frequency of 260 Hz, and the resulting current, $i_{\text{out}(t)}$, is amplified by a current amplifier (Femto DLPCA-200) connected to the
right reservoir electrode and measured using lock-in detection (Stanford Research Systems SR830).

\begin{figure}
    \includegraphics[scale=0.90]{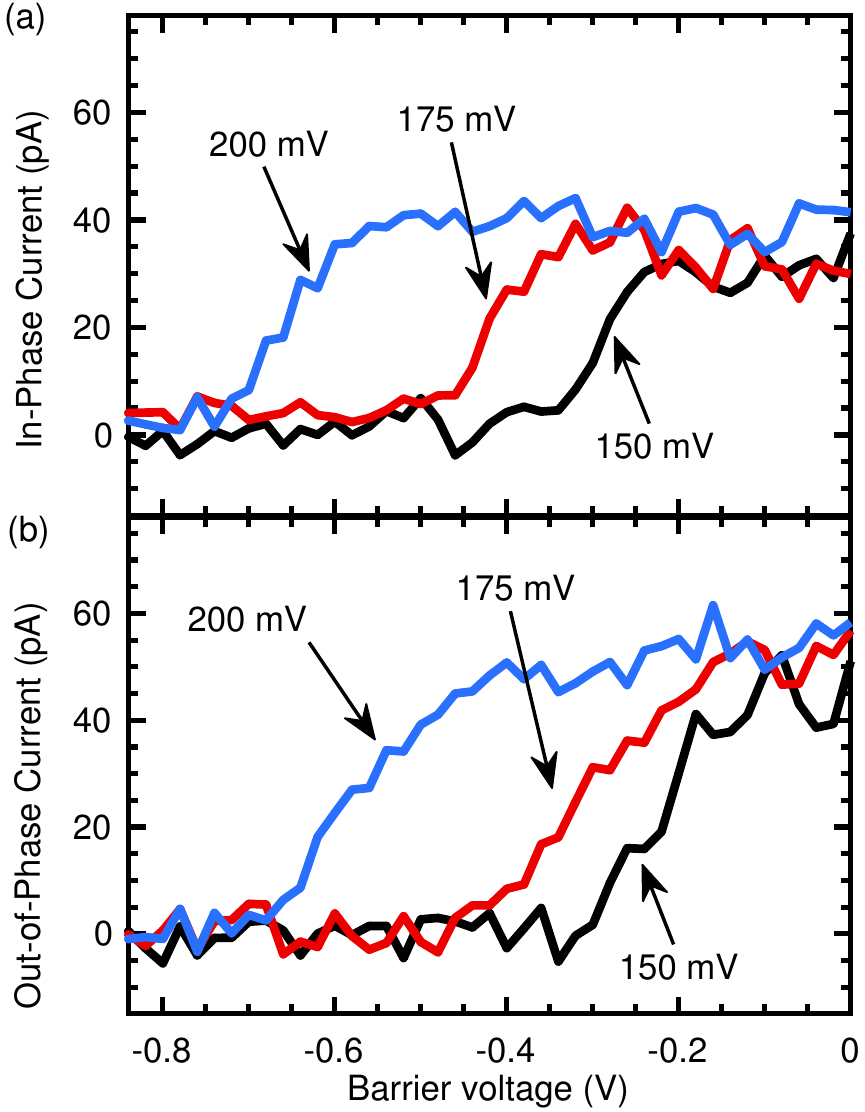}\caption{\label{fig3} Electron transport at three
    densities as a function of barrier voltage. Parts (a) and (b) show the in-phase and out-of-phase components of the
    signal measured from the right reservoir using lock-in detection. For electron densities of \{1.58, 1.97 and
    2.56\}\textdensity{}{9} set by ground plane voltages $V_\text{G} = \{-150,-175$ and $-200\}$ mV, the pinch-off
    barrier voltages are measured to be $V_\text{B} = \{-340,-480$ and $-720\}$ mV, respectively.
    }
\end{figure}

We begin by measuring transport across the device while sweeping the barrier voltage, $V_{\text{B}}$ more negative. Electrical transport
through the central microchannel can be controlled by the barrier voltage akin to a field-effect transistor. At low
barrier voltages, the central microchannel is open and allows current to flow between the two reservoirs. As the barrier
in the central microchannel is increased by sweeping the barrier voltage more negative, the resistance of the central
microchannel increases owing to the reduced density of electrons, as reflected by a reduction in the measured current.
This behavior continues until the potential at the barrier exceeds the electron potential in the microchannels and the
current is turned off completely. In \figref{fig3}a and b, we show the in-phase and out-of-phase current from the
lock-in measurement for three ground plane voltages of $V_\text{G} = \{-150,-175,-200\}$ mV indicating pinch-off at barrier voltages
$V_B = \{-340, -480, -720\}$ mV, respectively. These results can be interpreted by considering that more negative ground plane voltages
increase the density of electrons in the microchannels, which in turn require larger barrier potentials at the central
microchannel to cut off the transport. 

\begin{figure}
    \includegraphics{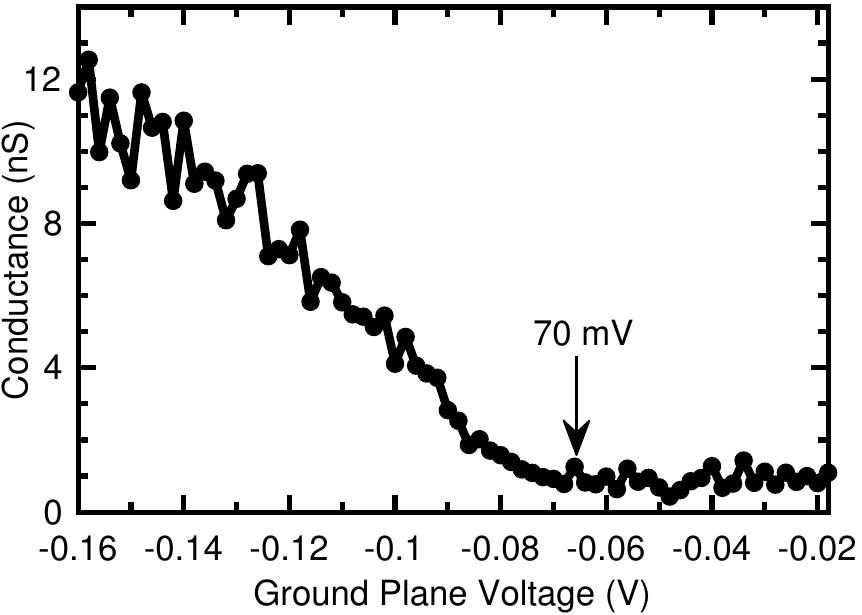}\caption{\label{fig2}Conductance of electron transport through the central
    microchannel. As the ground plane voltage exceeds $V_\text{G} = -70$ mV, the conductance is observed to increase up to a value of
    12 nS.}
\end{figure}

In order to estimate the density of electrons in the microchannels corresponding to the above ground plane voltages, we
open the barrier by setting it to 0 V and measure transport between the reservoirs as a function of the ground plane
voltage, $V_G$. The results are shown in \figref{fig2}.  For $V_G > -70$ mV, the channels above the left and right
reservoir electrodes are depleted, resulting in no measurable current through the central microchannel. On the other
hand, for $V_G < -70$ mV, the conductance is seen to increase up to a saturation value of 12 nS. The voltage sweep shown
in \figref{fig2} was repeated multiple times, with no appreciable hysteresis, indicating that electrons are readily
moved between the channels and the ground plane. In this case, the areal density, $n$, of electrons above the reservoir
electrodes can be determined from the relation $n = \frac{\epsilon_0\epsilon_{\text{He}}}{d}\times \left(V_G + 70\text{
    mV}\right)/e$, where $\epsilon_{\text{He}}$ is the dielectric constant of liquid Helium, $\epsilon_0$ is the
dielectric permittivity of free space, $e$ is the electronic charge, and $d=280$ nm is the thickness of the helium level
above the reservoir electrodes estimated from the height of the sample above the bulk helium level in our cell. From
this expression, we estimate the electron density above the microchannels to be \{1.58, 1.97, 2.56\}$\times10^9\text{
    cm}^{-2}$ for ground plane voltages of $V_\text{G} = \{-150, -175, -200\}$ mV, respectively. We note here that non-zero depletion
voltage has been observed in other microchannel devices\cite{rees_transport_2012}, and is likely due to a combination of
work function differences between the different metallic layers as well as thermal offset voltages.

\begin{figure}
    \includegraphics{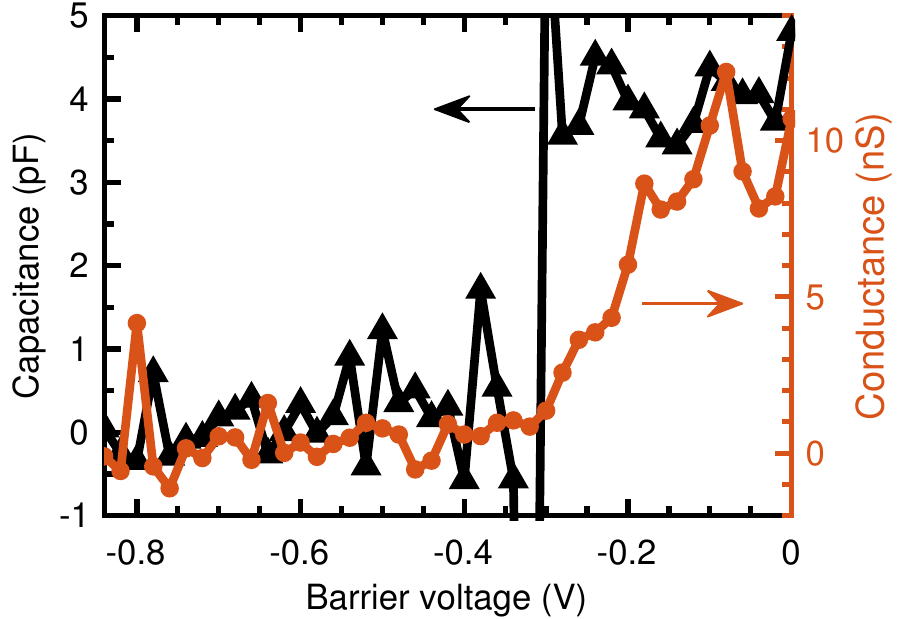}\caption{Capacitance above reservoirs and conductance through the central
    microchannel for an electron density of \textdensity{1.58}{9}. From the measured values, we estimate that the
    channels are half-filled above the reservoirs, and the mobility of electron transport is \textmobility{300}.\label{fig4}}
\end{figure}

In order to estimate the mobility of the electron transport, we analyze the data for the ground plane voltage set to
$V_\text{G} = -150$ mV in detail. Assuming that the device can be modeled as a series RC circuit ($R$ due to the central microchannel,
and $2C$ due to coupling to the 2D electron gas in the microchannels above each reservoir electrode), we extract $1/R$ and $C$ as
a function of barrier voltage. The results of these calculations are shown in \figref{fig4}. We find that the total
capacitance, $C$, remains nearly constant at a value of $\sim$4 pF once the barrier is opened, while the conductance
increases up to a saturation value of $\sim$10 nS. The mobility can then be estimated from the expression $\rho/\square
= 1/ne\mu$ where $\square \approx 8$ is the number of squares in the central microchannel, $\rho = (10\text{ nS})^{-1}$
is the 2D resistivity and $n = $\textdensity{1.58}{9} is the density of electrons in the microchannels for $V_G = -150$
mV. From this expression, we estimate the mobility of transport in the channels to be \textmobility{300}. In addition,
from the total area of the microchannels above each reservoir of $0.54$ mm$^2$, we estimate that the channels are
half-filled at this density.

We remark here on the assumptions made in the estimate of the mobility. First, we have assumed that the resistance $R$
of the transport comes entirely from the central microchannel. This assumption ignores the contribution of the $\sim900$
\textmicron\ long microchannels above the reservoirs to the resistance. Additionally, there is a 100 nm gap in the
central microchannel that has been used to separate the left and right reservoir electrodes. As an electron moves
between the reservoirs, it experiences a resistance due to the discontinuity seen by its image charge at the edges of
the gap. Given that the gap width of 100 nm is comparable to the distance of the electron above the reservoir electrodes
($d \approx 280$ nm), it is likely that the edge contributes to the resistance of the transport. Our
estimate of the mobility constitutes a lower bound as a result of these assumptions and further measurements are
necessary to separately measure these contributions to the resistance.

The ability to perform measurements with shallow helium has two advantages -- first, experiments with microchannel
devices that have previously been used where the helium is of order 1 \textmicron\ or thicker can now be extended to
shallow helium levels where larger electron densities can be supported. The second benefit is that devices that are
typically used in the study of solid-state quantum computation can be realized when gate electrodes can be patterned
beneath thin helium films. For example, these gate electrodes can be used to electrostatically define quantum dots
for individual electrons on helium, enabling the storage and manipulation of quantum information
\cite{dahm_quantum_2003,dykman_qubits_2003,lyon_spin-based_2006,schuster_proposal_2010}.

In conclusion, we have demonstrated transport of electrons on helium in a microchannel device where the channels are
only 200 nm deep. The choice of amorphous \fullTaWSi\ as the material for the fabrication of the channels has enabled
transport at these depths where the surface roughness and work function variations across the surface are minimized. We have
shown the ability to control the transport of electrons across reservoirs using a barrier as well as the ability
to set the density of electrons using the ground plane. Future work will investigate shallow devices where higher
electron densities can be supported. Additionally, the ability to work with shallow microchannels enables the design of
devices where quantum-mechanical effects such as lateral tunneling between quantum dots can be observed. 

\begin{acknowledgments}
Devices were fabricated in the Princeton Institute for the Science and Technology of Materials Micro/Nano Fabrication
    Laboratory and the Princeton University Quantum Device Nanofabrication Laboratory. Work at Princeton was supported
    by the NSF, in part through Grant No.\ DMR-1506862, and in part through the Princeton MRSEC (Grant No.\
    DMR-1420541). Sandia National Laboratories is a multimission laboratory managed and operated by National Technology
    \& Engineering Solutions of Sandia, LLC, a wholly owned subsidiary of Honeywell International Inc., for the U.S.
    Department of Energy's National Nuclear Security Administration under contract DE-NA0003525.  This paper describes
    objective technical results and analysis. Any subjective views or opinions that might be expressed in the paper do
    not necessarily represent the views of the U.S.  Department of Energy or the United States Government. 
\end{acknowledgments}


\begin{thebibliography}{19}%
\makeatletter
\providecommand \@ifxundefined [1]{%
 \@ifx{#1\undefined}
}%
\providecommand \@ifnum [1]{%
 \ifnum #1\expandafter \@firstoftwo
 \else \expandafter \@secondoftwo
 \fi
}%
\providecommand \@ifx [1]{%
 \ifx #1\expandafter \@firstoftwo
 \else \expandafter \@secondoftwo
 \fi
}%
\providecommand \natexlab [1]{#1}%
\providecommand \enquote  [1]{``#1''}%
\providecommand \bibnamefont  [1]{#1}%
\providecommand \bibfnamefont [1]{#1}%
\providecommand \citenamefont [1]{#1}%
\providecommand \href@noop [0]{\@secondoftwo}%
\providecommand \href [0]{\begingroup \@sanitize@url \@href}%
\providecommand \@href[1]{\@@startlink{#1}\@@href}%
\providecommand \@@href[1]{\endgroup#1\@@endlink}%
\providecommand \@sanitize@url [0]{\catcode `\\12\catcode `\$12\catcode
  `\&12\catcode `\#12\catcode `\^12\catcode `\_12\catcode `\%12\relax}%
\providecommand \@@startlink[1]{}%
\providecommand \@@endlink[0]{}%
\providecommand \url  [0]{\begingroup\@sanitize@url \@url }%
\providecommand \@url [1]{\endgroup\@href {#1}{\urlprefix }}%
\providecommand \urlprefix  [0]{URL }%
\providecommand \Eprint [0]{\href }%
\providecommand \doibase [0]{http://dx.doi.org/}%
\providecommand \selectlanguage [0]{\@gobble}%
\providecommand \bibinfo  [0]{\@secondoftwo}%
\providecommand \bibfield  [0]{\@secondoftwo}%
\providecommand \translation [1]{[#1]}%
\providecommand \BibitemOpen [0]{}%
\providecommand \bibitemStop [0]{}%
\providecommand \bibitemNoStop [0]{.\EOS\space}%
\providecommand \EOS [0]{\spacefactor3000\relax}%
\providecommand \BibitemShut  [1]{\csname bibitem#1\endcsname}%
\let\auto@bib@innerbib\@empty
\bibitem [{\citenamefont {Andrei}(1997)}]{andrei_two-dimensional_1997}%
  \BibitemOpen
  \bibfield  {author} {\bibinfo {author} {\bibfnamefont {E.~Y.}\ \bibnamefont
  {Andrei}},\ }\href@noop {} {\emph {\bibinfo {title} {Two-Dimensional Electron
  Systems on Helium and other Cryogenic Substrates}}}\ (\bibinfo  {publisher}
  {Springer Netherlands},\ \bibinfo {year} {1997})\BibitemShut {NoStop}%
\bibitem [{\citenamefont {Monarkha}\ and\ \citenamefont
  {Kono}(2004)}]{monarkha_two-dimensional_2004}%
  \BibitemOpen
  \bibfield  {author} {\bibinfo {author} {\bibfnamefont {Y.}~\bibnamefont
  {Monarkha}}\ and\ \bibinfo {author} {\bibfnamefont {K.}~\bibnamefont
  {Kono}},\ }\href@noop {} {\emph {\bibinfo {title} {Two-{Dimensional}
  {Coulomb} {Liquids} and {Solids}}}}\ (\bibinfo  {publisher}
  {Springer-Verlag},\ \bibinfo {year} {2004})\BibitemShut {NoStop}%
\bibitem [{\citenamefont {Peeters}\ and\ \citenamefont
  {Platzman}(1983)}]{peeters_electrons_1983}%
  \BibitemOpen
  \bibfield  {author} {\bibinfo {author} {\bibfnamefont {F.~M.}\ \bibnamefont
  {Peeters}}\ and\ \bibinfo {author} {\bibfnamefont {P.~M.}\ \bibnamefont
  {Platzman}},\ }\href {\doibase 10.1103/PhysRevLett.50.2021} {\bibfield
  {journal} {\bibinfo  {journal} {Phys. Rev. Lett.}\ }\textbf {\bibinfo
  {volume} {50}},\ \bibinfo {pages} {2021} (\bibinfo {year}
  {1983})}\BibitemShut {NoStop}%
\bibitem [{\citenamefont {Jiang}\ \emph {et~al.}(1988)\citenamefont {Jiang},
  \citenamefont {Stan},\ and\ \citenamefont {Dahm}}]{jiang_phase_1988}%
  \BibitemOpen
  \bibfield  {author} {\bibinfo {author} {\bibfnamefont {H.~W.}\ \bibnamefont
  {Jiang}}, \bibinfo {author} {\bibfnamefont {M.~A.}\ \bibnamefont {Stan}}, \
  and\ \bibinfo {author} {\bibfnamefont {A.~J.}\ \bibnamefont {Dahm}},\ }\href
  {\doibase 10.1016/0039-6028(88)90655-3} {\bibfield  {journal} {\bibinfo
  {journal} {Surface Science}\ }\textbf {\bibinfo {volume} {196}},\ \bibinfo
  {pages} {1} (\bibinfo {year} {1988})}\BibitemShut {NoStop}%
\bibitem [{\citenamefont {Etz}\ \emph {et~al.}(1984)\citenamefont {Etz},
  \citenamefont {Gombert}, \citenamefont {Idstein},\ and\ \citenamefont
  {Leiderer}}]{etz_stability_1984}%
  \BibitemOpen
  \bibfield  {author} {\bibinfo {author} {\bibfnamefont {H.}~\bibnamefont
  {Etz}}, \bibinfo {author} {\bibfnamefont {W.}~\bibnamefont {Gombert}},
  \bibinfo {author} {\bibfnamefont {W.}~\bibnamefont {Idstein}}, \ and\
  \bibinfo {author} {\bibfnamefont {P.}~\bibnamefont {Leiderer}},\ }\href
  {\doibase 10.1103/PhysRevLett.53.2567} {\bibfield  {journal} {\bibinfo
  {journal} {Phys. Rev. Lett.}\ }\textbf {\bibinfo {volume} {53}},\ \bibinfo
  {pages} {2567} (\bibinfo {year} {1984})}\BibitemShut {NoStop}%
\bibitem [{\citenamefont {Hu}\ and\ \citenamefont
  {Dahm}(1990)}]{hu_stability_1990}%
  \BibitemOpen
  \bibfield  {author} {\bibinfo {author} {\bibfnamefont {X.~L.}\ \bibnamefont
  {Hu}}\ and\ \bibinfo {author} {\bibfnamefont {A.~J.}\ \bibnamefont {Dahm}},\
  }\href {\doibase 10.1103/PhysRevB.42.2010} {\bibfield  {journal} {\bibinfo
  {journal} {Phys. Rev. B}\ }\textbf {\bibinfo {volume} {42}},\ \bibinfo
  {pages} {2010} (\bibinfo {year} {1990})}\BibitemShut {NoStop}%
\bibitem [{\citenamefont {Grimes}\ and\ \citenamefont
  {Adams}(1980)}]{grimes_crystallization_1980}%
  \BibitemOpen
  \bibfield  {author} {\bibinfo {author} {\bibfnamefont {C.~C.}\ \bibnamefont
  {Grimes}}\ and\ \bibinfo {author} {\bibfnamefont {G.}~\bibnamefont {Adams}},\
  }\href {\doibase 10.1016/0039-6028(80)90465-3} {\bibfield  {journal}
  {\bibinfo  {journal} {Surface Science}\ }\textbf {\bibinfo {volume} {98}},\
  \bibinfo {pages} {1} (\bibinfo {year} {1980})}\BibitemShut {NoStop}%
\bibitem [{\citenamefont {Mistura}\ \emph {et~al.}(1997)\citenamefont
  {Mistura}, \citenamefont {Günzler}, \citenamefont {Neser},\ and\
  \citenamefont {Leiderer}}]{mistura_microwave_1997}%
  \BibitemOpen
  \bibfield  {author} {\bibinfo {author} {\bibfnamefont {G.}~\bibnamefont
  {Mistura}}, \bibinfo {author} {\bibfnamefont {T.}~\bibnamefont {Günzler}},
  \bibinfo {author} {\bibfnamefont {S.}~\bibnamefont {Neser}}, \ and\ \bibinfo
  {author} {\bibfnamefont {P.}~\bibnamefont {Leiderer}},\ }\href {\doibase
  10.1103/PhysRevB.56.8360} {\bibfield  {journal} {\bibinfo  {journal} {Phys.
  Rev. B}\ }\textbf {\bibinfo {volume} {56}},\ \bibinfo {pages} {8360}
  (\bibinfo {year} {1997})}\BibitemShut {NoStop}%
\bibitem [{\citenamefont {Angrik}\ \emph {et~al.}(2004)\citenamefont {Angrik},
  \citenamefont {Faustein}, \citenamefont {Klier},\ and\ \citenamefont
  {Leiderer}}]{angrik_electrons_2004}%
  \BibitemOpen
  \bibfield  {author} {\bibinfo {author} {\bibfnamefont {J.}~\bibnamefont
  {Angrik}}, \bibinfo {author} {\bibfnamefont {A.}~\bibnamefont {Faustein}},
  \bibinfo {author} {\bibfnamefont {J.}~\bibnamefont {Klier}}, \ and\ \bibinfo
  {author} {\bibfnamefont {P.}~\bibnamefont {Leiderer}},\ }\href {\doibase
  10.1023/B:JOLT.0000049060.15686.b6} {\bibfield  {journal} {\bibinfo
  {journal} {Journal of Low Temperature Physics}\ }\textbf {\bibinfo {volume}
  {137}},\ \bibinfo {pages} {335} (\bibinfo {year} {2004})}\BibitemShut
  {NoStop}%
\bibitem [{\citenamefont {Klier}\ \emph {et~al.}(2008)\citenamefont {Klier},
  \citenamefont {Doicescu}, \citenamefont {Leiderer},\ and\ \citenamefont
  {Shikin}}]{klier_electron_2008}%
  \BibitemOpen
  \bibfield  {author} {\bibinfo {author} {\bibfnamefont {J.}~\bibnamefont
  {Klier}}, \bibinfo {author} {\bibfnamefont {I.}~\bibnamefont {Doicescu}},
  \bibinfo {author} {\bibfnamefont {P.}~\bibnamefont {Leiderer}}, \ and\
  \bibinfo {author} {\bibfnamefont {V.}~\bibnamefont {Shikin}},\ }\href
  {\doibase 10.1007/s10909-007-9537-0} {\bibfield  {journal} {\bibinfo
  {journal} {J Low Temp Phys}\ }\textbf {\bibinfo {volume} {150}},\ \bibinfo
  {pages} {212} (\bibinfo {year} {2008})}\BibitemShut {NoStop}%
\bibitem [{\citenamefont {McGlone}(2017)}]{mcglonethesis}%
  \BibitemOpen
  \bibfield  {author} {\bibinfo {author} {\bibfnamefont {J.~M.}\ \bibnamefont
  {McGlone}},\ }\emph {\bibinfo {title} {Development of Amorphous Metal Thin
  Films for Thermal Inkjet Printing and Microelectronics}},\ \href@noop {}
  {Ph.D. thesis},\ \bibinfo  {school} {Oregon State University} (\bibinfo
  {year} {2017})\BibitemShut {NoStop}%
\bibitem [{\citenamefont {McGlone}\ \emph {et~al.}(2015)\citenamefont
  {McGlone}, \citenamefont {Olsen}, \citenamefont {Stickle}, \citenamefont
  {Abbott}, \citenamefont {Pugliese}, \citenamefont {Long}, \citenamefont
  {Keszler},\ and\ \citenamefont {Wager}}]{mcglone_ta-based_2015}%
  \BibitemOpen
  \bibfield  {author} {\bibinfo {author} {\bibfnamefont {J.~M.}\ \bibnamefont
  {McGlone}}, \bibinfo {author} {\bibfnamefont {K.~R.}\ \bibnamefont {Olsen}},
  \bibinfo {author} {\bibfnamefont {W.~F.}\ \bibnamefont {Stickle}}, \bibinfo
  {author} {\bibfnamefont {J.~E.}\ \bibnamefont {Abbott}}, \bibinfo {author}
  {\bibfnamefont {R.~A.}\ \bibnamefont {Pugliese}}, \bibinfo {author}
  {\bibfnamefont {G.~S.}\ \bibnamefont {Long}}, \bibinfo {author}
  {\bibfnamefont {D.~A.}\ \bibnamefont {Keszler}}, \ and\ \bibinfo {author}
  {\bibfnamefont {J.~F.}\ \bibnamefont {Wager}},\ }\href {\doibase
  10.1016/j.jallcom.2015.07.226} {\bibfield  {journal} {\bibinfo  {journal}
  {Journal of Alloys and Compounds}\ }\textbf {\bibinfo {volume} {650}},\
  \bibinfo {pages} {102} (\bibinfo {year} {2015})}\BibitemShut {NoStop}%
\bibitem [{\citenamefont {McGlone}\ \emph {et~al.}(2017)\citenamefont
  {McGlone}, \citenamefont {Olsen}, \citenamefont {Stickle}, \citenamefont
  {Abbott}, \citenamefont {Pugliese}, \citenamefont {Long}, \citenamefont
  {Keszler},\ and\ \citenamefont {Wager}}]{mcglone_tawsi_2017}%
  \BibitemOpen
  \bibfield  {author} {\bibinfo {author} {\bibfnamefont {J.~M.}\ \bibnamefont
  {McGlone}}, \bibinfo {author} {\bibfnamefont {K.~R.}\ \bibnamefont {Olsen}},
  \bibinfo {author} {\bibfnamefont {W.~F.}\ \bibnamefont {Stickle}}, \bibinfo
  {author} {\bibfnamefont {J.~E.}\ \bibnamefont {Abbott}}, \bibinfo {author}
  {\bibfnamefont {R.~A.}\ \bibnamefont {Pugliese}}, \bibinfo {author}
  {\bibfnamefont {G.~S.}\ \bibnamefont {Long}}, \bibinfo {author}
  {\bibfnamefont {D.~A.}\ \bibnamefont {Keszler}}, \ and\ \bibinfo {author}
  {\bibfnamefont {J.~F.}\ \bibnamefont {Wager}},\ }\href {\doibase
  10.1557/mrc.2017.77} {\bibfield  {journal} {\bibinfo  {journal} {MRS
  Communications}\ }\textbf {\bibinfo {volume} {7}},\ \bibinfo {pages} {715}
  (\bibinfo {year} {2017})}\BibitemShut {NoStop}%
\bibitem [{\citenamefont {Yang}\ \emph {et~al.}(2016)\citenamefont {Yang},
  \citenamefont {Fragner}, \citenamefont {Koolstra}, \citenamefont {Ocola},
  \citenamefont {Czaplewski}, \citenamefont {Schoelkopf},\ and\ \citenamefont
  {Schuster}}]{yang_coupling_2016}%
  \BibitemOpen
  \bibfield  {author} {\bibinfo {author} {\bibfnamefont {G.}~\bibnamefont
  {Yang}}, \bibinfo {author} {\bibfnamefont {A.}~\bibnamefont {Fragner}},
  \bibinfo {author} {\bibfnamefont {G.}~\bibnamefont {Koolstra}}, \bibinfo
  {author} {\bibfnamefont {L.}~\bibnamefont {Ocola}}, \bibinfo {author}
  {\bibfnamefont {D.}~\bibnamefont {Czaplewski}}, \bibinfo {author}
  {\bibfnamefont {R.}~\bibnamefont {Schoelkopf}}, \ and\ \bibinfo {author}
  {\bibfnamefont {D.}~\bibnamefont {Schuster}},\ }\href {\doibase
  10.1103/PhysRevX.6.011031} {\bibfield  {journal} {\bibinfo  {journal} {Phys.
  Rev. X}\ }\textbf {\bibinfo {volume} {6}},\ \bibinfo {pages} {011031}
  (\bibinfo {year} {2016})}\BibitemShut {NoStop}%
\bibitem [{\citenamefont {Rees}\ \emph {et~al.}(2012)\citenamefont {Rees},
  \citenamefont {Kuroda}, \citenamefont {Marrache-Kikuchi}, \citenamefont
  {Höfer}, \citenamefont {Leiderer},\ and\ \citenamefont
  {Kono}}]{rees_transport_2012}%
  \BibitemOpen
  \bibfield  {author} {\bibinfo {author} {\bibfnamefont {D.~G.}\ \bibnamefont
  {Rees}}, \bibinfo {author} {\bibfnamefont {I.}~\bibnamefont {Kuroda}},
  \bibinfo {author} {\bibfnamefont {C.~A.}\ \bibnamefont {Marrache-Kikuchi}},
  \bibinfo {author} {\bibfnamefont {M.}~\bibnamefont {Höfer}}, \bibinfo
  {author} {\bibfnamefont {P.}~\bibnamefont {Leiderer}}, \ and\ \bibinfo
  {author} {\bibfnamefont {K.}~\bibnamefont {Kono}},\ }\href {\doibase
  10.1007/s10909-011-0416-3} {\bibfield  {journal} {\bibinfo  {journal} {J Low
  Temp Phys}\ }\textbf {\bibinfo {volume} {166}},\ \bibinfo {pages} {107}
  (\bibinfo {year} {2012})}\BibitemShut {NoStop}%
\bibitem [{\citenamefont {Dahm}(2003)}]{dahm_quantum_2003}%
  \BibitemOpen
  \bibfield  {author} {\bibinfo {author} {\bibfnamefont {A.~J.}\ \bibnamefont
  {Dahm}},\ }\href {\doibase 10.1063/1.1542534} {\bibfield  {journal} {\bibinfo
   {journal} {Low Temperature Physics}\ }\textbf {\bibinfo {volume} {29}},\
  \bibinfo {pages} {489} (\bibinfo {year} {2003})}\BibitemShut {NoStop}%
\bibitem [{\citenamefont {Dykman}\ \emph {et~al.}(2003)\citenamefont {Dykman},
  \citenamefont {Platzman},\ and\ \citenamefont
  {Seddighrad}}]{dykman_qubits_2003}%
  \BibitemOpen
  \bibfield  {author} {\bibinfo {author} {\bibfnamefont {M.~I.}\ \bibnamefont
  {Dykman}}, \bibinfo {author} {\bibfnamefont {P.~M.}\ \bibnamefont
  {Platzman}}, \ and\ \bibinfo {author} {\bibfnamefont {P.}~\bibnamefont
  {Seddighrad}},\ }\href {\doibase 10.1103/PhysRevB.67.155402} {\bibfield
  {journal} {\bibinfo  {journal} {Phys. Rev. B}\ }\textbf {\bibinfo {volume}
  {67}},\ \bibinfo {pages} {155402} (\bibinfo {year} {2003})}\BibitemShut
  {NoStop}%
\bibitem [{\citenamefont {Lyon}(2006)}]{lyon_spin-based_2006}%
  \BibitemOpen
  \bibfield  {author} {\bibinfo {author} {\bibfnamefont {S.~A.}\ \bibnamefont
  {Lyon}},\ }\href {\doibase 10.1103/PhysRevA.74.052338} {\bibfield  {journal}
  {\bibinfo  {journal} {Phys. Rev. A}\ }\textbf {\bibinfo {volume} {74}},\
  \bibinfo {pages} {052338} (\bibinfo {year} {2006})}\BibitemShut {NoStop}%
\bibitem [{\citenamefont {Schuster}\ \emph {et~al.}(2010)\citenamefont
  {Schuster}, \citenamefont {Fragner}, \citenamefont {Dykman}, \citenamefont
  {Lyon},\ and\ \citenamefont {Schoelkopf}}]{schuster_proposal_2010}%
  \BibitemOpen
  \bibfield  {author} {\bibinfo {author} {\bibfnamefont {D.~I.}\ \bibnamefont
  {Schuster}}, \bibinfo {author} {\bibfnamefont {A.}~\bibnamefont {Fragner}},
  \bibinfo {author} {\bibfnamefont {M.~I.}\ \bibnamefont {Dykman}}, \bibinfo
  {author} {\bibfnamefont {S.~A.}\ \bibnamefont {Lyon}}, \ and\ \bibinfo
  {author} {\bibfnamefont {R.~J.}\ \bibnamefont {Schoelkopf}},\ }\href
  {\doibase 10.1103/PhysRevLett.105.040503} {\bibfield  {journal} {\bibinfo
  {journal} {Phys. Rev. Lett.}\ }\textbf {\bibinfo {volume} {105}},\ \bibinfo
  {pages} {040503} (\bibinfo {year} {2010})}\BibitemShut {NoStop}%
\end{thebibliography}
\end{document}